\documentclass[pra,twocolumn,showpacs]{revtex4-1}
\usepackage{ae}
\usepackage{graphicx}
\usepackage{amssymb}
\usepackage{amsmath}
\usepackage{amsthm}
\usepackage{color}
\usepackage{sidecap}
\usepackage{ulem}

\newcommand{\rb}{$^{87}$Rb}

\newcommand{\kq}{$^{41}$K}

\newcommand{\fref}[1]{Fig.~\ref{#1}}

\newcommand{\mr}{\mathrm}

\newcommand{\gkrb}{g_\mr{1D,KRb}}
\newcommand{\grb}{g_\mr{1D,Rb}}
\newcommand{\idest}{{\it i.e.}}

\begin{document}
\title{Quantum dynamics of impurities in a 1D Bose gas}

\author{J.~Catani$^{1,2}$}
\author{G.~Lamporesi$^{1,2}$}
\author{D.~Naik$^{1}$}
\author{M.~Gring$^{3}$}
 \author{M.~Inguscio$^{1,2}$}
 \author{F.~Minardi$^{1,2}$}
\email{francesco.minardi@ino.it}
\affiliation{$^1$LENS-European Laboratory for Non-Linear
  Spectroscopy and Dipartimento di Fisica, Universit\`a di Firenze,
  via N. Carrara 1, IT-50019 Sesto Fiorentino-Firenze, Italy\\
  $^2$CNR-INO, via G. Sansone 1, IT-50019 Sesto Fiorentino-Firenze, Italy\\
$^3$Vienna Center for Quantum Science and Technology, Atominstitut,
  TU-Wien, AT-1020 Vienna, Austria}
\author{A.~Kantian, T.~Giamarchi}
\affiliation{DPMC-MaNEP, University of Geneva, 24 Quai Ernest-Ansermet, CH-1211
  Geneva, Switzerland}

\begin{abstract}
  Using a species-selective dipole potential, we create
  initially localized impurities and investigate their interactions
  with a majority species of bosonic atoms in a one-dimensional 
  configuration during expansion. We find an interaction-dependent
  amplitude reduction of the oscillation of the impurities' size with
  no measurable frequency shift, and study it as a function of the
  interaction strength.  We discuss possible theoretical interpretations 
  of the data. We compare, in particular, with a 
  polaronic mass shift model derived following Feynman variational approach.
\end{abstract}


\date{\today}

\maketitle

\section{Introduction}
Low-dimensional and strongly interacting systems have sparked intense
scientific interest in recent years in diverse research
fields, such as solid-state physics, nanoscience and atomic physics. In
particular, strongly correlated systems display quantum phases
dominated by quantum fluctuations, such as the Mott-insulator phase
\cite{Imada_mott_RMP1998} and magneticlike ordered phases
\cite{Auerbach_book_spins}. In low dimensions, the interplay between
interactions among particles and confining potential can enhance the
effect of quantum correlations, resulting in peculiar regimes, such as
Tonks-Girardeau or the sine-Gordon \cite{Giamarchi_Book1D} regime. Due to the
unprecedented control over the interatomic interactions, the external
trapping potentials, and the internal states of the atoms, ultracold
atomic systems represent a versatile tool to explore these novel
phenomena, and have already provided the way to realize some of these
quantum phases \cite{Bloch_cold_atoms_optical_lattices_RMP2008}. A
particular strength of cold atoms is the realization of systems that
are hard to obtain in condensed-matter physics, such as
multicomponent bosons of either different hyperfine states
\cite{Myatt_SpinorCondensate_PRL1997,Stenger_SpinorCondensate_Nature1998}
or different species \cite{Modugno_RbKBEC_PRL2002}, as well
as the study of real-time dynamics of quantum many-body systems.

\begin{figure}[b]
\centering
\includegraphics[width=0.95\columnwidth]{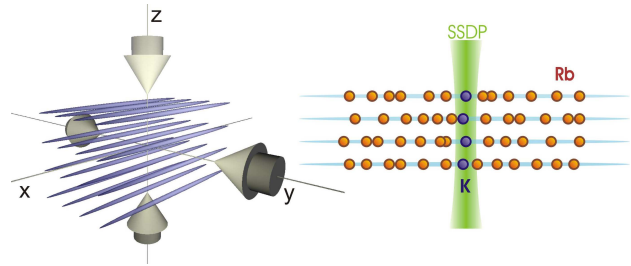}
\caption{(Color online) Ultracold Rb atoms with K impurities are
  loaded into an array of 1D systems, ``tubes'' (left). The SSDP
  light blade spatially localizes the impurities into the center of
  the Rb tubes (right). }
\label{fig:sketch}
\end{figure}

In particular, cold atoms provide a way to realize a recently proposed
new universal class of quantum systems, the so-called ferromagnetic
liquids \cite{Zvonarev_ferrobosons_PRL2007}, occurring as the ground
state of repulsively interacting two-component bosons. The
time evolution of a spin flip, \idest, the excitation obtained by
flipping one single bosonic ``spin'' from the fully spin-polarized
ground state, thereby creating a localized ``impurity'', leads to a
class of dynamics completely different to that of the standard
(Luttinger-liquid) theory but possessing remarkable universal features
\cite{Zvonarev_ferrobosons_PRL2007,
  Imambekov_structfactor_LiebLiniger_Bethe08,Zvonarev_YangGaudin_PRB2009}.
The physics of such impurities propagating in a sea of the majority
species is further connected to several longstanding questions in
strongly correlated systems, both fermionic and bosonic. For fermionic
systems and an immobile impurity, one example is the well-known x-ray
edge problem, with the impurity leading to the Anderson orthogonality
catastrophe\cite{Mahan_ManyParticlePhysics}. The motion of the
impurity then strongly affects this physics, as was probed both for
impurities in $^3$He \cite{Rosch_heavy_particle_fermions_PRL1988} and,
more recently, for minority species in interacting fermionic systems
\cite{Schirotzek_FermionicPolaron_PRL2009,
  Nascimbene_FermionicPolaron_PRL2009}, in the context of Fermi-liquid
theory. Inextricably linked to the study of the impurity motion is the
notion of polarons \cite{Feynman_statmech}, \idest, the occurrence of
density fluctuations of the majority species, both for fermions and
bosons, leading to renormalization of the impurity parameters, such as
its mass. In one dimension, however, the nature of the bath is quite
special, leading to the above-mentioned novel physical effects and, in
particular, to subdiffusion at zero temperature. This problem is also
directly related to the motion of a driven impurity which was
investigated recently both theoretically
\cite{Astrakharchik_driven_impurity_luttinger_PRA2004,
  Cherny_decay_sf_currents_PRA2009,
  Gangardt_Bloch_oscillations_1D_spinor_PRL2009} and experimentally
\cite{Palzer_Impurity_thru_Tonks_PRL2009}. A system where the
impurity-bath interactions are adjustable over a wide range paves the
way for the study of a wealth of physical phenomena.

In this work, we realize such a system using a species-selective
dipole potential (SSDP) \cite{Catani_Entropy_PRL2009}. A minority
species (K atoms) can diffuse into a majority species of (Rb) bosonic
atoms, in a one-dimensional (1D) configuration (see
\fref{fig:sketch}). We study how the interspecies interactions
reduce the oscillation amplitude of the impurities' size
$\sigma(t)=\sqrt{\langle x^2\rangle}$. We compare the data in the
light of a model based on polaronic mass shifts.

The remainder of the paper is organized as follows: in Sec. II we
describe the experimental procedure to prepare the one-dimensional
(1D) systems of K impurities initially localized in the surrounding Rb
bath and measure their subsequent expansion; in Sec. III we
report the experimental data; Section IV illustrates our theoretical
model based on a quantum Langevin equation. By means of the Feynman
variational approach \cite{Feynman_Polarons_1955}, we calculate the
impurities' mass renormalization due to the interactions with the Rb
bath.  In Sec. V we compare the data with the results of the
theoretical model, discussing the limitations and approximations
involved. Finally, we summarize in Sec. VI.

\section{Experimental procedure}

The mixture of \rb\ and \kq\ is first cooled to 1.5\,$\mu$K by
microwave evaporation of Rb and sympathetic cooling of K in a magnetic
trap, then loaded into a crossed dipole trap created by two orthogonal
laser beams ($\lambda$=1064\,nm, waists$\,\simeq 70\,\mu$m). Both
species are spin polarized in their $|F$=$1, m_F$=$1\rangle$ hyperfine
states, featuring magnetically tunable interspecies interactions in
the $0-10$ mT range. The mixture is further cooled by optical
evaporation performed in a uniform magnetic field of 7.73 mT to
adjust the interspecies scattering length $a$ to the convenient value
of $240$ Bohr radii, ensuring both fast thermalization and a low rate
of inelastic collisions.  

At this point we set the magnetic field to 7.15 mT, corresponding to
vanishing interspecies interactions, and we adiabatically raise a
vertical ($z$) 1D optical lattice (waist 170 $\mu$m) to
15 (6.5)\,$E_\mr{rec}$ for Rb (K) in 200 ms. The lattice wavelength
$\lambda$=1064 nm results in a 2.3 times smaller lattice height for
\kq\ than for \rb\, in units of the respective recoil energies
($E_\mr{rec}=h^2/2m\lambda^2$).  After adiabatic extinction of the
dipole trap, both species move downward as collisions disrupt the
gravity-induced Bloch oscillations \cite{Ott_PRL2004} but, due to
largely different tunneling times, K drops faster and, by adjusting
the fall time, we reduce the gravitational sag between the two species
in the dipole trap.

\begin{figure}[t!]
\centering
\includegraphics[width=0.95\columnwidth]{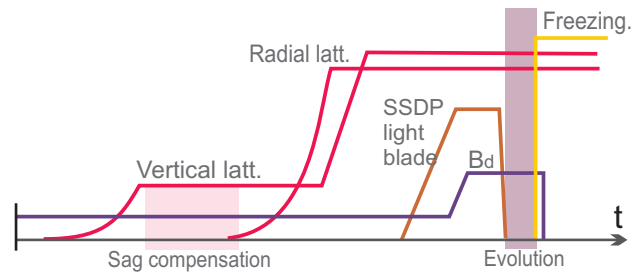}
\caption{(Color online) Time sequence of the experimental procedure
  used to prepare the K impurities in the middle of 1D tubes filled
  with Rb atoms.  }
\label{fig:time_sequence}
\end{figure}

Subsequently, we further raise the vertical lattice to
$60\,E_\mr{rec}$ for Rb and adiabatically switch on an additional
standing wave with equal waist and strength along the $y$
direction. Thus we create the array of 1D systems (tubes) by means of
a two-dimensional (2D) lattice. The lattice transverse harmonic oscillator frequency,
$\omega_\bot/2\pi = $34 (45) kHz for Rb(K), exceeds both the temperature
and the chemical potential, thus ensuring the 1D regime for both
species. 

The residual trapping frequency along the direction $x$ of the tubes,
62 (87) Hz for Rb(K), is due to the inhomogeneous transverse profile
of the lattice beams. An SSDP elliptic beam,
orthogonal to the tubes, is then turned on in 50\,ms, compressing K at
the center of Rb and leaving the latter nearly unperturbed.  This {\it
  light blade} has waists of 15 and 75\ $\mu$m ($x$ and $z$ direction,
respectively), wavelength of 770.4 nm, and power equal to
0.6 mW. Correspondingly, the depth and frequency are 11 $\mu$K and
$\omega_K/(2\pi)=1.0$ kHz.

Finally, the interaction strength $\gkrb$ between K and Rb, is brought
to the desired value by linearly ramping the magnetic field to its
final value $B_d$. Instead, the interaction strength of the Rb atoms
$\grb=2.36\times10^{-37}$ J$\,$m is independent of the applied
magnetic field $B_d$.  

The impurities' dynamics is initiated by rapidly extinguishing the
light blade with a linear ramp of 0.5 ms.  Images of the impurities
are taken once their motion has been frozen by suddenly adding a tight
optical lattice along the tubes axis allowing for magnetic field to be
extinguished in 15 ms and for atoms to be repumped in the hyperfine
level suitable for imaging without significantly affecting their
density distribution. The full experimental procedure is shown in
\fref{fig:time_sequence}.

Typically, we measure $(1.8\pm 0.2)\times10^5$ \rb\ and $(6\pm
2)\times10^3$ \kq\ atoms at 140 nK before loading the tubes. From the
analysis of time-of-flight images, we obtain a temperature of the Rb
sample in the 1D tubes of 350(50) nK. We estimate the Rb peak
filling and density to be 180 atoms/tube and
$n_\mr{1D,Rb}\simeq7 \mu$m$^{-1}$, with a filling-averaged
Lieb-Liniger parameter $(m_{Rb} \grb)/(\hbar^2 n_\mr{1D,Rb}) \simeq 1$; the peak K filling is approximately 1.4 atom/tube.

\section{Experimental data}

\begin{figure*}[t!]
\centering
\includegraphics[width=1.95\columnwidth]{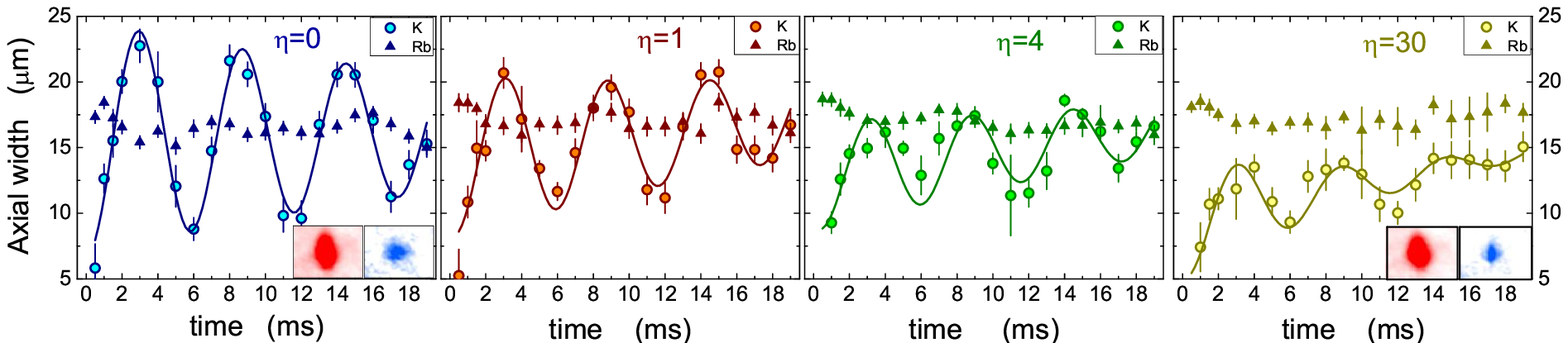}

\includegraphics[width=1.7\columnwidth]{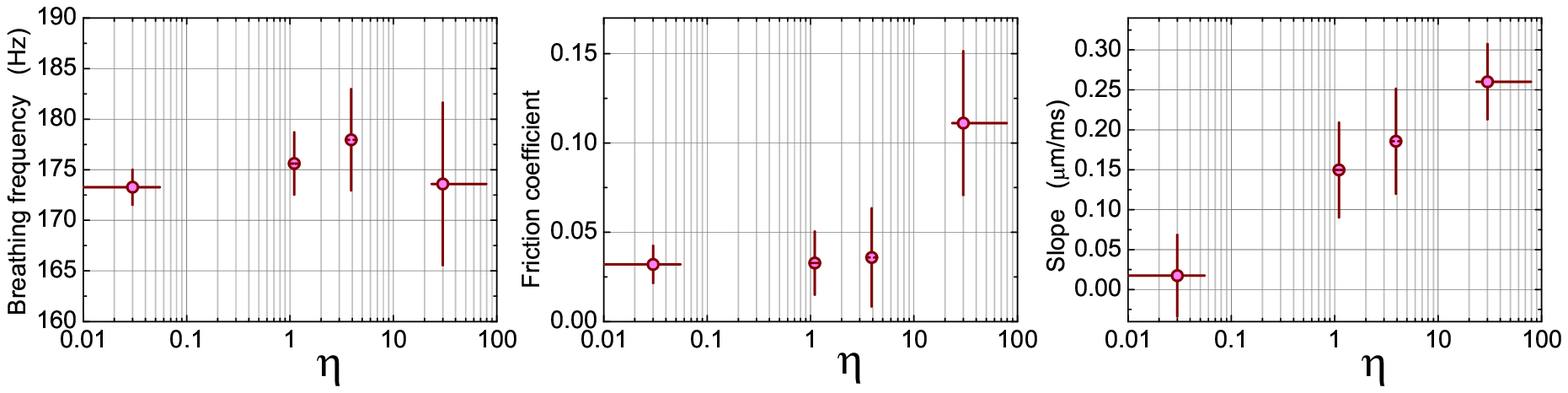}

\caption{(Color online) Upper panel: oscillations of the K impurities
  axial width $\sigma=\sqrt{\langle x^2 \rangle}$ after extinction of
  the confining SSDP, for different interaction strengths with the
  surrounding Rb bulk, $\eta=\gkrb/\grb$. Solid lines represent a fit
  to data, triangles the Rb axial size. Inset images show the observed
  in-situ density distributions of Rb (red, left) and K (blue, right)
  within a window of $200 \mu$m $\times 150 \mu$m. Lower panel: values
  of the fit parameters $\omega/2\pi$, $\gamma$ and $\beta$ (left to
  right, see text for definition).}
\label{fig:oscillations}
\end{figure*}

We record the oscillation of the impurities axial size
$\sigma(t)=\sqrt{\langle x^2\rangle}$ along the tubes as a function of
time, through resonant {\it in situ} absorption imaging. We
fit the absorption profiles by a two-dimensional Gaussian
function. Figure \ref{fig:oscillations} shows the evolution of the impurities axial
size $\sigma(t)$ for four different values of the 1D interspecies
interaction strength $\gkrb$. Given the magnetic field $B_d$, hence
the interspecies scattering length $a$
\cite{Thalhammer_Tunable_double_bec_PRL2008}, $\gkrb$ is calculated
following \cite{Peano_Confined_hetero_scattering_NJP2005} and
expressed in units of $\grb$ as $\gkrb =\eta  \grb$.  We find the
oscillation amplitude of $\sigma(t)$ to be large for $\eta\simeq 0$
(no interactions between impurities and 1D gas) and to decrease with
$\eta$. Remarkably, for the largest oscillation strength, the
impurities oscillations are always confined inside the bath. In
addition, we observe that, in addition to oscillating, $\sigma(t)$ increases
linearly (at least initially) over time.
In order to extract quantitative information, we fit the experimental
data with a damped sine function with linear baseline:
$\sigma(t)$=$\sigma_1+\beta t-A\,e^{-\gamma \omega
  t}\,\cos(\sqrt{1-\gamma^2}\omega\,(t-t_0))$. Since in the following
we will mainly focus on the oscillation amplitude, in
\fref{fig:oscillations} we show the fit parameters, $\omega/2\pi,\,
\gamma,\, \beta$.

We first notice that the oscillation frequency does not shift, within
our error bars, with $\gkrb$: such a surprising feature is further
discussed below. Then we observe that, even in the absence of
interactions at $\eta=0$, a residual damping occurs, likely due
to intraspecies collisions in tubes with more than one K atom. We also
notice that, during the oscillations, the impurity cloud can grow larger
than the Rb sample, thus exploring the Rb inhomogeneous density
profile. Such inhomogeneity along the tubes axis adds to the
inhomogeneity of the tube's filling. In the following, we account for
the inhomogeneity using the local density approximation, \idest, by
considering homogeneous systems with different values of Rb linear
density. As K impurities are initially confined in the middle of Rb,
the first oscillation is less affected by Rb inhomogeneous density
than the following. Therefore we focus our analysis on the value of
the maximum size reached in the first oscillation, after 
3 ms of expansion: $\sigma_p\equiv \sigma(t=3$ ms). Such value
closely reflects the oscillation amplitude $A$ as, for short times,
both the slope and the damping have negligible effect.

\begin{figure}[b!]
\includegraphics[width=\columnwidth]{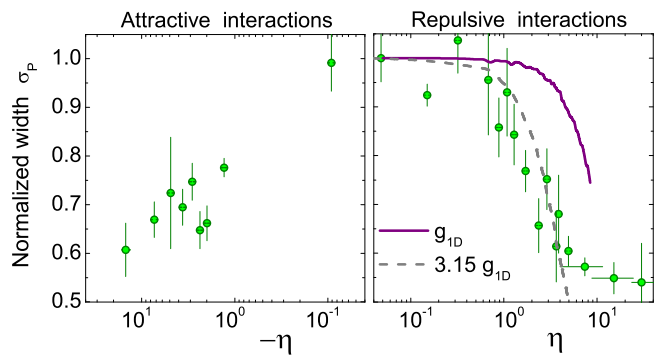}
\caption{(Color online) Experiment: impurities' axial size at the
  first oscillation maximum $\sigma_p$, normalized so that
  $\sigma_p=1$ for $\eta=0$, versus the coupling strength parameter
  $\eta=\gkrb/\grb$, for attractive (left, $\eta<0$) and repulsive
  (right, $\eta>0$) interactions (circles). Theory: $\sqrt{m_K/m_K^*}$
  calculated with impurity-bath coupling $\gkrb$ derived from two-body
  scattering \cite{Peano_Confined_hetero_scattering_NJP2005} (solid
  line) and with $\gkrb$ multiplied by a fit parameter (dashed line). Both
  curves are computed for a Rb density of 7 atoms/$\mu$m and
  $\tilde{\gamma}= 0$.}
\label{fig:norm_sigma}
\end{figure}

Figure \ref{fig:norm_sigma} shows the dependence of $\sigma_p$ on the
relative coupling strength $\eta$: values of $\eta$ of the order of
the unity are sufficient to induce a significant amplitude reduction,
if compared to the non interacting case $\eta=0$.  It is important to
remark that, if mean-field interactions dominated, we would observe an
opposite behavior between positive and negative $\eta$ values. As
$\sigma_p$ decreases with $|\eta|$, independently of its sign, we
conclude that mere mean-field interactions cannot explain our
experimental findings. We also notice a saturation behavior for
$\eta>4$ that cannot be due to the washing out of the Feshbach
resonance caused by magnetic field instability ($\sim
10\,\mu$T). Instead, it might be related to a crossover to the
three-dimensional (3D) regime, as we estimate that for $\eta=15$ the
mean-field interaction energy of the K impurities, $\gkrb n_{{\rm
    1D,Rb}}$, equals the band-gap of the 2D lattice. As our theory
cannot describe this 1D to 3D crossover, and furthermore the inclusion
of the 1D bound states between K and Rb for $\eta<0$ also cannot be
adequately captured, we have calculated $\sqrt{m_K/m_K^*}$ only in the
range $0<\eta\leq 8$ in Fig. 3.

\begin{figure}[t!]
\centering

\includegraphics[width=\columnwidth]{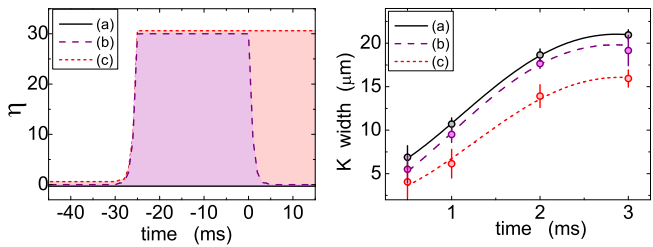}

\caption{(Color online) Amplitude of the first oscillation of the
  axial size of K impurities (right) following different preparation
  sequences (left): (a) $\eta$ always negligible; (b) $\eta$ large
  while K in the SSDP light blade and negligible during expansion at
  $t>0$; and (c) $\eta$ always large.}
\label{fig:thermalization}
\end{figure}

To some extent, the amplitude of the first oscillation could reflect
different preparation temperatures of the K sample in the light blade,
as the efficiency of the thermalization with the Rb background depends
on the strength of interactions.  Thus, we recorded the first
oscillation of K impurities prepared at large $\eta$ and expanding at
zero interactions, and compared with the oscillations of samples
prepared and expanded at large (or zero) interactions. We notice from
\fref{fig:thermalization} that the preparation has a nearly negligible impact
on the oscillation amplitude, insufficient to explain the large
differences observed between high and small $\eta$. Moreover,
incomplete thermalization cannot explain why at large interactions
$\sigma_p$ lies below the Rb axial width, as shown in
\fref{fig:oscillations}.

To summarize, the key experimental findings are (a) even for large
$\eta$ the oscillation frequency of $\sigma(t)$ does not deviate
measurably from that of the noninteracting case (which is $2\omega$);
(b) the larger $|\eta|$, the lower the amplitude of oscillation
initially;
(c) in the interacting case, besides
oscillating, the impurities width increases over time in a seemingly
linear fashion; and (d) at long times $\sigma(t)$ equilibrates to about the
same value, independently of $\eta$.

\section{Theoretical model}
A complete explanation of the observed phenomena is an interesting and
open problem. Motivated by the observed shape of $\sigma(t)$, we give
here a semiempirical analysis of the oscillation through the model of
a damped, quantum harmonic oscillator in contact with a thermal bath,
\idest, a quantum Langevin equation \cite{Gardiner_QuantumNoise},
$\dot{\hat{x}}(t)=\hat{p}(t)/m_K^*$, $\dot{\hat{p}}(t)=-k^*\hat{x}(t)-
\tilde\gamma\hat{p}(t)+\hat{\xi}(t)$, as a framework to order and
provide a first interpretation of the findings. Here, $\hat{x}(t)$ and
$\hat{p}(t)$ are respectively the position and momentum operators of
one \kq\ atom whose mass $m_K^*$ and spring constant $k^*$ are
renormalized by interactions with the bath. In addition to the
exponential damping $-\tilde{\gamma}\hat{p}(t)$, we account for the
fluctuations of the bath by the noise operator $\hat{\xi}(t)$, whose
correlator $\langle\hat{\xi}(t)\hat{\xi}(t')\rangle \propto
\tilde{\gamma}$ is approximated by using the thermal quantum
statistics of a set of harmonic oscillators at constant spectral
density \cite{Gardiner_QuantumNoise}. The operator $\hat{\xi}(t)$
causes the gradual increase of the impurities' width, corresponding to
heating from the thermal component of the bath.

Various effects could account for the change of amplitude. A dominant
mean-field interaction for $\eta>0$ would make the effective potential
for the K shallower, resulting in $\sigma_p$ \textit{increasing} with
$\eta$, at odds with observed behavior. Another possibility is the
damping from the $\tilde\gamma$ term. However this is also
inconsistent with the observed data: within the quantum Langevin
approximation, and at the experimental temperatures, any but very
small values of $\tilde\gamma$ result in strong heating contributions
from the correlator $\langle\hat{\xi}(t)\hat{\xi}(t')\rangle$, 
% to $\sigma_p$ that cancel out any decrease from the exponential damping ,
such that overall $\sigma_p$ increases with $\eta$.

At variance with the two above-mentioned effects, a mass
renormalization of the K impurities due to polaronic effects
\cite{Feynman_Polarons_1955} is a very good starting point to explain
the decrease of $\sigma_p$. 

\subsection{Polaronic mass shift}
We calculate the mass shift $M=m_K^*-m_K$ using Feynman's variational
theory for the polaron \cite{Feynman_Polarons_1955}.
Taking the combined impurity-bath Hamiltonian
\begin{equation}
  \hat{H}=\frac{\hat{p}^2}{2m_K}+\sum_{k\neq 0}\epsilon_k
  \hat{b}_k^{\dagger}\hat{b}_k+\sum_{k\neq 0}V_ke^{ik\hat{x}}\left(b_k+b_{-k}^{\dagger}\right),
\end{equation}
where the Rb bath has been approximated as a Tomonaga-Luttinger liquid
with linearized density fluctuations around a homogeneous background
density, described by the bosonic operators $\hat{b}_k$
$\hat{b}_k^{\dagger}$ with $\epsilon_k=v_s|k|$,
$V_k=\gkrb\left(K|k|/2\pi L\right)^{1/2}e^{-|k|/2k_c}$, and where
$v_s$ is the sound velocity of Rb, $K$ the Luttinger parameter,
related to the effective Rb-Rb interaction and $k_c$ the cutoff
momentum.  These quantities can be related to a homogeneous 1D Rb gas
with arbitrary contact interactions \cite{CazalillaReview_RMP2011}; as a
working theory for inhomogeneous 1D systems is difficult to obtain, we
use this homogeneous theory instead, evaluating it for different Rb
densities in the center of the trap. We find that the results are not
that strongly depending on the precise value of the Rb density.

\begin{figure}[t!]
\centering
\includegraphics[width=0.95\columnwidth]{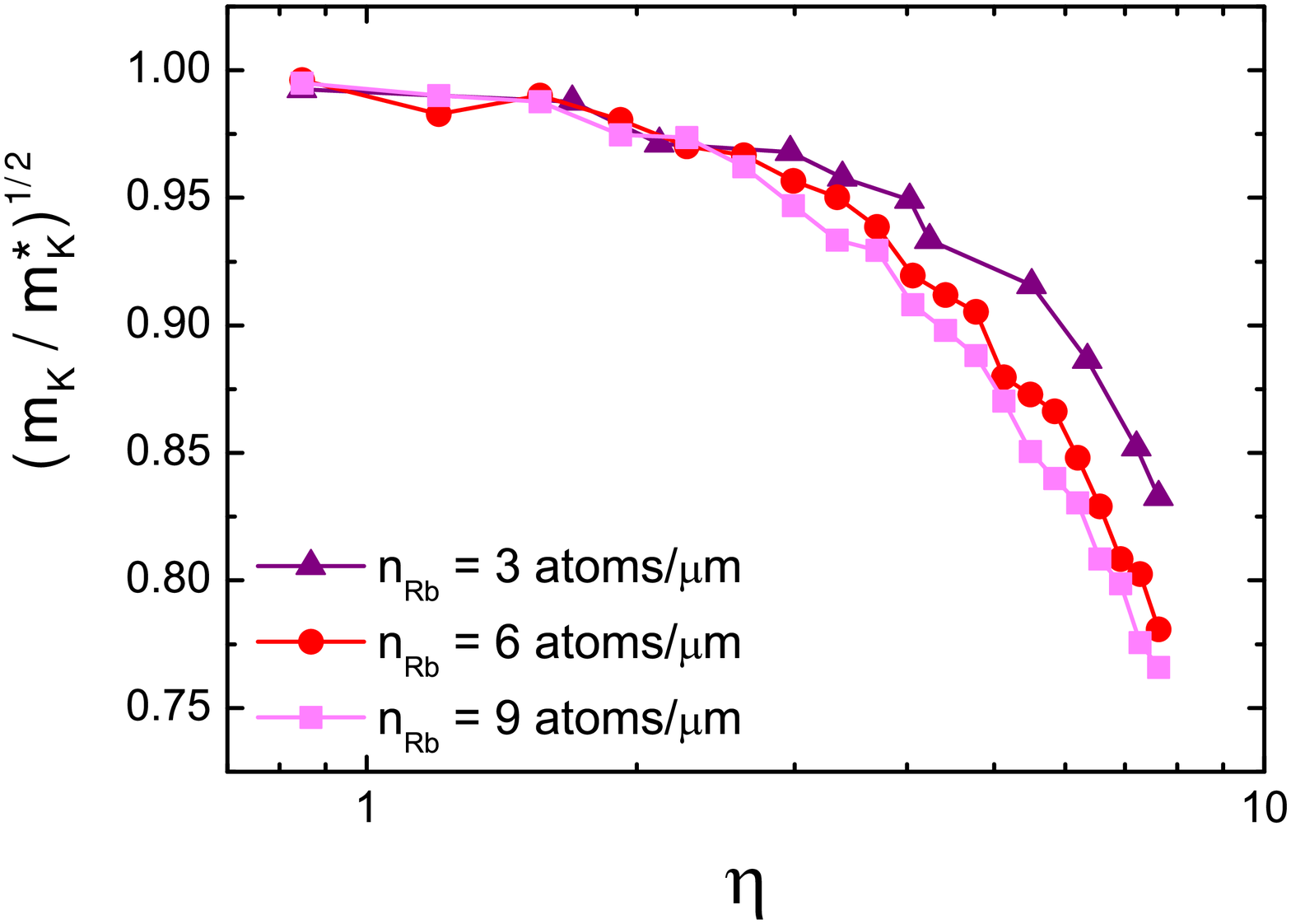}
\caption{(Color online) Square root of the inverse impurities' renormalized 
mass versus interspecies interaction strength $\eta$, calculated for different 
  values of Rb density. This quantity directly compares with the measured K width $\sigma_p$.
%All other parameters are as for the theory in
%  \fref{fig:norm_sigma}.
}
\label{fig:densvar}
\end{figure}

Evaluating the full partition function $Z=\mr{tr}\left(e^{-\beta
    \hat{H}}\right)$ as a path integral and integrating out the Rb
bath yields an action $S$ for the K atom that is non-local in time:

\begin{eqnarray} \label{eq:action}
  &&S=\int_0^{\beta\hbar}d\tau \frac{m_K}{2}\dot{x}^2(\tau) - \\
  &&\sum_k
  \frac{V_k^2}{2\hbar}\int_0^{\beta\hbar}d\tau\int_0^{\beta\hbar}d\tau'
  G(k,|\tau-\tau'|) e^{ik[x(\tau)-x(\tau')]}, \nonumber
\end{eqnarray}

where
$G(u,k)=\cosh(\epsilon_k(u-\hbar\beta/2))/\sinh(\hbar\beta\epsilon_k/2)$
The mass shift enters through the Feynman trial action $S_0$,
\begin{eqnarray*} 
    && S_0=\int_0^{\beta\hbar}d\tau
    \frac{m_K}{2}\dot{x}^2(\tau) + \frac{MW^3}{8} \\
&&\times\int\int_0^{\beta\hbar}d\tau d\tau'
    \frac{\cosh(W|\tau-\tau'|-W\hbar\beta/2)}{\sinh(W\beta\hbar/2)}[x(\tau)-x(\tau')]^2
\end{eqnarray*}
which depends on two parameters, the mass shift $M$ and $W$, which
are chosen by minimizing the trial free energy
$F_0+\langle S-S_0\rangle /(\hbar\beta)\geq F= - (1/\beta)\ln Z$.

The explicit expression of the trial free energy is

\begin{eqnarray}
   && F_0+\frac{1}{\beta \hbar}\langle S-S_0\rangle \nonumber\\
&&= \ln\left[\sinh\left(\frac{\beta\hbar W\alpha}{2}\right)\right] - \ln\left[\sinh\left(\frac{\beta\hbar W}{2}\right)\right]-\ln\alpha \nonumber \\
   & & -\frac{M}{2(m_K+M)}\left[\frac{\hbar\beta
        W\alpha}{2}\coth\left(\frac{\beta\hbar
          W\alpha}{2}\right)-1\right] \nonumber \\
&&-\sum_k
    \frac{V_k^2}{\hbar}\int_0^{\hbar\beta}du\left(1+\frac{u}{\hbar\beta}\right)G(k,u)\mathcal{K}(k,u),
\end{eqnarray}

where $\alpha=\sqrt{1+M/m_K}$, and

\begin{eqnarray}
\mathcal{K}&=&
\exp\left[ -\frac{\hbar k^2}{2(m_K+M)}
\left(u-\frac{u^2}{\hbar\beta} + \frac{M}{m_K} \right. \right. \nonumber \\
&&=\left. \left. \times \frac{\cosh(W\alpha\hbar\beta/2)-\cosh[W\alpha(\hbar\beta/2-u)]}{W\alpha\sinh(\hbar\beta W\alpha/2)} \right)\right],
\end{eqnarray}

which we then minimize numerically to obtain the mass shift $M$.

We apply this theory to an experiment with an inhomogeneous density
distribution of bath particles (due to the external confinement of the
Rb atoms). For this approach to be reasonable, the mass shift should
not depend too strongly on a shift in Rb density, which
\fref{fig:densvar} shows to be indeed the case (all other parameters
are as in the experiment).

Note that one can easily add a term corresponding to a parabolic
confining potential in the action (\ref{eq:action}) and repeat the
variational procedure to find the effective mass. Quite intuitively,
one finds that the tighter the confinement, the smaller is the mass
renormalization.  This strongly suggests that before the blade is
released one should consider that the K impurity has its bare
mass. This is clearly a point that would need to be more
quantitatively tested by, e.g., a Monte Carlo calculation.

\section{Comparison of data and theory}

Strongly simplifying the bath (see \cite{Gardiner_QuantumNoise}) for a
small $\tilde\gamma$ as the data suggest we obtain the functional
form for $\sigma(t)$ used to fit the oscillations in
\fref{fig:oscillations}. The theoretically predicted decrease of
$\sigma_p$ is proportional to  $1/\sqrt{m_K^*}$. Therefore, in
\fref{fig:norm_sigma} we compare the measured
$\sigma_p$ normalized to the value at zero interactions to the calculated values of 
$\sqrt{m_K/m_K^*}$: the theoretical curve reproduces the experimental
trend and magnitude of the amplitude reduction fairly well.
%, as shown in 
%, where we have used Feynmans' variational
%ansatz for the polaron mass shift \cite{Feynman_Polarons_1955} to
%calculate $m_K^*$, and where we assume $\tilde{\gamma}= 0$. 

As the
variational approach cannot take the inhomogeneity of the bath into
account, we approximate the translationally invariant theory with
representative values for the Rb density in the center of the trap. We
find that, in the parameter range of the experiment, the mass shift is
rather insensitive to the Rb density. 

We also considered the possibility that $\gkrb$ in the experiment
(\idest, at finite density) may be different from the one obtained in
the standard 1D two-body scattering theory \cite{Olshanii_CIR_PRL1998,
  Peano_Confined_hetero_scattering_NJP2005}.  While those formulas
have indeed been demonstrated to well describe the position of the
confined-induced resonances (CIR) also in the regime of finite
density, it is currently unknown how accurate it is for the actual
value of $\gkrb$ away from the CIR. We find that a least square fit of
the theory to data with a value of $3.15\gkrb$ fits the experiment
better.  

Concerning the mass shift theory for $\sigma_p$, we must note
that for the trap parameters and temperatures it relies on the
assumption that the mass of K is not or is only insignificantly
renormalized inside the light blade during preparation. This is
supported by calculations of the mass renormalization that include a
parabolic trap for the K, which show the mass shift becoming smaller
the deeper the trap, eventually becoming zero. This simple extension
of Feynman's approach is, however, ill suited to quantitative
calculations of concrete values of the mass shift in a tight
confinement, and we therefore must leave the quantitative test of this
assumption to strong numerical methods.

The above interpretation based on the polaronic mass shift also
predicts an increase of the oscillation period, which for large $\eta$
would be beyond the $5\%$ experimental error bars.  To reconcile this
apparent contradiction with the data, we need to consider the physical
processes resulting in an effective upward renormalization of the
spring constant $k^*$. Such processes are more easily visualized in
the limit of impenetrability: when the impurities cannot get across
the bath atoms, their displacement from the trap bottom costs the
additional harmonic potential energy of the bath atoms moving uphill.
While this is observed in proof-of-principle simulations for
simplified models, quantitatively estimating the value of $k^*$ is
difficult with current methods and will require further, more refined
analysis.

\section{Conclusions}
In conclusion, we have investigated the dynamics of impurities in 1D
atomic samples as we varied their interaction strength with the
surrounding bath of weakly interacting bosons by more than two orders
of magnitude. We observe the amplitude of the impurities' quadrupole
oscillations decreasing with the absolute value of the interaction
strength, a fact unexplained by mere mean-field interactions. An
analysis of the data in the light of a simplified quantum Langevin
equation suggests that polaronic mass shift plays a role in the
amplitude reduction. 

\section{Acknowledgments}
This work was supported by MIUR PRIN 2007, Ente CdR Firenze, EU under
STREP NAME-QUAM and CHIMONO, ERC DISQUA and the Swiss FNS under MaNEP
and Division II. M. G. is supported by Austrian Science Fund (FWF):
W1210, CoQuS.

\bibliography{./Diffusion1D}

\end{document}